\documentclass[aps,prl,preprint,amssymb]{revtex4}
\usepackage{graphicx}
\begin{document}
\title{Using a laser to cool a semiconductor}
\author{Danhong Huang\footnote{danhong.huang@kirtland.af.mil}, T. Apostolova, P. M. Alsing, and D. A. Cardimona}
\affiliation{Air Force Research Lab, Space Vehicles Directorate,
Kirtland Air Force Base, New Mexico 87117}
\date{\today}

\begin{abstract}
A nonlocal energy-balance equation is derived for the optical
absorption, photoluminescence and inelastic electron-phonon
scattering, which determines the electron and hole temperatures
for any given lattice temperature. The evolution of the lattice
temperature is found to be determined by the difference between
the power-loss density due to photoluminescence and the power-gain
density due to optical absorption, as well as by the initial
lattice temperature. We find that in addition to the expected
decrease in the lattice temperature, the electron temperature also
decreases with time. A laser-cooling power as high as $380$ eV$/$s
is predicted for the wide bandgap semiconductor AlN initially at
room temperature when the pump-laser field is only $10$ V$/$cm.
Laser cooling is found to be more efficient for a large bandgap
material, a weaker laser field, and a high initial lattice
temperature. The laser-cooling rate then decreases as the lattice
cools. The theory presented here provides quantitative predictions
that can guide future experiments.
\end{abstract}
\maketitle

The cooling of a solid via light-induced fluoresce has been of
interest for a very long time
\cite{pring,gauck,gang,edwards,fernandez}. This interesting
phenomenon involves the excitation of an electron from the valence
bandedge to the conduction bandedge by absorbing a pump photon.
This cool electron quickly becomes hot by gaining thermal energy
through ultrafast electron-phonon scattering. After a radiative
lifetime, recombination of the hot electron will produce a
spontaneous photon with energy higher than that of the pump
photon. As a result, the lattice will be cooled due to the loss of
thermal energy to the electron. It is only recently that this
phenomenon has been observed experimentally. Laser-induced
fluorescent cooling of heavy-metal-fluoride glass doped with
trivalent ytterbium ions was the first realization of this concept
\cite{richard}. Soon to follow were demonstrations of cooling in
dye solutions \cite{clark} and thulium-doped glass \cite{hoyt}.
\medskip

Laser cooling of a semiconductor, however, remains an elusive
goal, although it has been pursued for many years
\cite{oraevsky,rivlin}. Indeed, now more than ever the field of
laser cooling is a topic of intense on-going theoretical and
experimental investigations \cite{muri}. The only theories
attempting to model the laser cooling phenomenon are local
simulation theories that include rate equations for determining
the steady-state carrier density and the loss of lattice energy
with several kinetic coefficients. The problem with these theories
is that they neglect important dynamical effects such as the
change of the carrier distribution when the temperature is
lowered. Therefore, they only apply to situations with little
change of temperature. The main feature of the rate equation
approach is its simplicity, but it is unable to elucidate the
essential physics behind the laser cooling phenomena. The key
question that still remains open is, what are the best
semiconductor materials and conditions for achieving the greatest
laser cooling effect? This requires an accurate {\em nonlocal}
theory on a microscopic level, which directly provides an
evolution equation for the lattice temperature by including the
dynamical effects. This theory should establish a criterion for
the occurrence of laser cooling. The theory should also establish
a criterion for the efficiency of the laser cooling if it does
exist. In this paper, we will present such a {\em nonlocal} theory
for the laser cooling of semiconductors. By including the effect
of the carrier distribution, we will be able to uncover the
essential physics underlying the phenomenon and we will be able to
provide important quantitative predictions that can guide
experimentalists toward achieving maximum efficiency of laser
cooling in the future.
\medskip

Let us first recall the transport force-balance theory for hot
electron transport \cite{lei0}. When a {\em dc} electric field is
applied to a doped semiconductor, there is a drift of electrons as
a result of the center-of-mass motion. This is described by a
balancing of forces between the frictional forces due to
scattering processes and the electrostatic force. At the same
time, electrons form a quasi-equilibrium Fermi-Dirac distribution.
The electron temperature of this distribution can be determined by
the energy balance between the power-gain density due to joule
heating and the power-loss density due to electron-phonon
scattering. As a result, the electron temperature becomes higher
than the lattice temperature if the lattice is in thermal
equilibrium with a heat bath and the drift velocity is large.
\medskip

For the situation under consideration in this paper, a weak pump
laser first excites electrons from the valence bandedge to the
conduction bandedge. The excited carriers instantaneously form a
non-equilibrium distribution \cite{rink}. It is well known that
the quantum kinetics of the scattering of electrons with phonons
or other carriers under a weak pump field can only be seen within
the time scale of several hundred femtoseconds \cite{koch1}.
Subsequently, ultrafast carrier-phonon and carrier-carrier
scattering quickly adjusts the kinetic energies of these excited
carriers by taking energy from the lattice \cite{koch}. As a
result, a quasi-equilibrium Fermi-Dirac distribution of carriers
is formed in about $0.1\ ps$ \cite{rink}, with an electron
temperature determined by the pump-field intensity, pump-photon
energy, and lattice temperature. After a few tens of nanoseconds,
radiative decay of the excited carriers will begin to affect the
electron distribution. The electron temperature will be
adiabatically readjusted according to an energy balance between
the power-gain density due to optical absorption, the power-loss
density due to photoluminescence, and the power-exchange density
due to scattering with phonons. At the same time, the lattice
temperature will evolve because of an imbalance between the power
loss due to transferring phonon energy to carriers and the slight
power gain from the external thermal radiation. Just before the
radiative decay occurs, the lattice and the electrons are in
thermal equilibrium with an initial temperature which can be
determined by solving a semiconductor Bloch equation \cite{koch}.
\medskip

Using the nonlocal theory described below, we find that the
laser-cooling rate is largest for a large bandgap material, a
weaker pump-laser field, and a high initial lattice temperature.
We also find that the laser-cooling power decreases as the lattice
cools down.
\medskip

In general, the power gain by electrons from the absorption of the
pump laser can not be balanced by the power loss due to
spontaneous photon emission alone. As a result, electrons either
take energy from or give energy to phonons through inelastic
scattering, which depends on the sign of the difference between
the electron and lattice temperatures. The electron temperature
can be determined by an energy-balance equation for any given
lattice temperature. The pair scattering between electrons due to
the Coulomb interaction conserves the total energy and does not
contribute to the energy-balance equation. On the other hand, the
single-particle electron-phonon scattering greatly contributes to
the energy-balance equation.
\medskip

The power-density loss due to spontaneous photon emission is
calculated to be \cite{lyo,dave}

\[
{\cal W}_{\rm sp}=\frac{\sqrt{\epsilon_{\rm
r}}e^2}{\pi\hbar^2m_0\epsilon_0c^3{\cal V}}\
\left(\frac{m_0}{m_{\rm e}^{\ast}}-1\right)\ \frac{E_{\rm
G}(E_{\rm G}+\Delta_0)}{E_{\rm G}+2\Delta_0/3}\ \sum_{\vec{k}}\
\left[E^{\prime}_{\rm
G}(k)+\frac{\hbar^2k^2}{2\mu^{\ast}}\right]^2\ f_k^{\rm e}f_k^{\rm
h}
\]
\begin{equation}
=\frac{\sqrt{\epsilon_{\rm r}}e^2}{2\pi^3\hbar^2m_0\epsilon_0c^3}\
\left(\frac{m_0}{m_{\rm e}^{\ast}}-1\right)\ \frac{E_{\rm
G}(E_{\rm G}+\Delta_0)}{E_{\rm G}+2\Delta_0/3}\ \int_0^{\infty}\
dk\ k^2\ \left[E^{\prime}_{\rm
G}(k)+\frac{\hbar^2k^2}{2\mu^{\ast}}\right]^2\ f_k^{\rm e}f_k^{\rm
h}\ , \label{e1}
\end{equation}
where $f_k^{\rm e}$ ($f_k^{\rm h}$) is the quasi-equilibrium
electron (hole) distribution at electron (hole) temperature
$T_{\rm e}$ ($T_{\rm h}$), $E_{\rm G}$ is the bandgap, $\Delta_0$
is the spin-orbit splitting, $m^{\ast}_{\rm e}$ is the electron
effective mass, $\mu^{\ast}$ is the reduced mass of electrons and
holes, $\epsilon_{\rm r}$ is the average relative dielectric
constant, ${\cal V}$ is the volume of the sample, and
$E^{\prime}_{\rm G}(k)$ is the renormalized bandgap. It is clear
from Eq.\,(\ref{e1}) that the larger the bandgap or the higher the
carrier temperature, the stronger the power-loss density will be.
\medskip

The power-density gain due to pumping by a spatially-uniform laser
for $\hbar\Omega_{\rm p}\ge E_{\rm G}$ is found to be \cite{huang}

\[
{\cal W}_{\rm p}=\frac{4}{\hbar{\cal V}}\ \sum_{\vec{k}}\
\Delta^2_k\left[E^{\prime}_{\rm
G}(k)+\frac{\hbar^2k^2}{2\mu^{\ast}}\right]\ \left(1-f_k^{\rm
e}-f_k^{\rm h}\right)\ \frac{\gamma_0}{\gamma_0^2+[E^{\prime}_{\rm
G}(k)+\hbar^2k^2/2\mu^{\ast}-\hbar\Omega_{\rm p}]^2}
\]
\[
=\frac{2}{\pi^2\hbar}\ \int_0^{\infty}\ dk\ k^2\
\Delta^2_k\left[E^{\prime}_{\rm
G}(k)+\frac{\hbar^2k^2}{2\mu^{\ast}}\right]\ \left(1-f_k^{\rm
e}-f_k^{\rm h}\right)
\]
\begin{equation}
\times\frac{\gamma_0}{\gamma_0^2+[E^{\prime}_{\rm
G}(k)+\hbar^2k^2/2\mu^{\ast}-\hbar\Omega_{\rm p}]^2}\ , \label{e2}
\end{equation}
where $\gamma_0$ is the homogeneous level broadening,
$\hbar\Omega_p$ is the pump-photon energy, and $2\Delta_k$ is the
renormalized Rabi splitting which is proportional to the square
root of the pump-laser intensity. It is seen from Eq.\,(\ref{e2})
that the greater the pump-laser field, the higher the power-gain
density will be. By comparing Eqs.\,(\ref{e1}) and (\ref{e2}), we
find that the ratio of the power-loss density to the power-gain
density scales as $E_{\rm G}^3/{\cal E}^2_{\rm p}$. For
simplicity, we do not include recapture of photoluminescence
photons here, which can be equivalently included as an adjustment
of the pump-laser intensity if the carrier temperature is much
smaller than $E_{\rm G}/k_{\rm B}$.
\medskip

By keeping only the leading order interaction between electrons
and phonons or impurities in the Heisenberg equation, we get the
following power-exchange density ${\cal W}^{\rm e}_{\rm s}$ from
impurities, phonons and scattering-assisted photons to electrons
\cite{kaiser,lei}

\[
{\cal W}^{\rm e}_{\rm s}=2\pi n_{\rm i}\ \sum_{\vec{q}}\
\left|U^{\rm e}_{\rm imp}(q)\right|^2\ \sum_{n=-\infty}^{\infty}\
\left[J_n\left(\frac{eq_{\|}{\cal E}_{\rm p}}{m_{\rm
e}^{\ast}\Omega^2_{\rm p}}\right)\right]^2\ n\Omega_{\rm p}
\]
\[
\times\sum_{\vec{k}}\ \left(f^{\rm e}_k-f^{\rm
e}_{|\vec{k}+\vec{q}|}\right)\ \delta\left(E^{\rm
e}_{|\vec{k}+\vec{q}|}-E^{\rm e}_k-n\hbar\Omega_{\rm p}\right)
\]
\[
-\frac{4\pi}{{\cal V}}\ \sum_{\vec{q},\lambda}\ \left|C^{\rm
e}_{q\lambda}\right|^2\ \sum_{n=-\infty}^{\infty}\
\left[J_n\left(\frac{eq_{\|}{\cal E}_{\rm p}}{m_{\rm
e}^{\ast}\Omega^2_{\rm p}}\right)\right]^2
\left(\omega_{q\lambda}-n\Omega_{\rm p}\right)\left[N^{\rm
ph}_0\left(\frac{\hbar\omega_{q\lambda}}{k_{\rm B}T_{\rm
L}}\right)-N^{\rm
ph}_0\left(\frac{\hbar\omega_{q\lambda}-n\hbar\Omega_{\rm
p}}{k_{\rm B}T_{\rm e}}\right)\right]
\]
\begin{equation}
\times\sum_{\vec{k}}\ \left(f^{\rm e}_k-f^{\rm
e}_{|\vec{k}+\vec{q}|}\right)\ \delta\left(E^{\rm
e}_{|\vec{k}+\vec{q}|}-E^{\rm
e}_k+\hbar\omega_{q\lambda}-n\hbar\Omega_{\rm p}\right)\ ,
\label{e3}
\end{equation}
which includes the phonon- or impurity-assisted photon absorption
for $n\ne 0$. Here, ${\cal E}_p$ is the pump-laser strength,
$E_k^{\rm e}$ is the renormalized electron kinetic energy, and
$T_{\rm e}$ and $T_{\rm L}$ are the electron and lattice
temperatures, $q_{\|}$ lies in the polarization direction of the
pump-laser field, $N_0^{\rm ph}(x)=[\exp(x)-1]^{-1}$ is the
Bose-Einstein function, $n$ is an integer, $J_n(x)$ is the $nth$
order Bessel function, $\hbar\omega_{q\lambda}$ is the phonon
energy for phonon wave number $q$ and mode $\lambda$, $n_{\rm i}$
is the impurity concentration, $|U^{\rm e}_{\rm
imp}(q)|^2=|e^2/[\epsilon_0\epsilon_{\rm r}(q^2+\Lambda^2_{\rm
e}){\cal V}]|^2$ is the electron-impurity coupling strength,
$1/\Lambda_{\rm e}$ is the static screening length, and $|C^{\rm
e}_{q\lambda}|^2$ is the electron-phonon coupling strength. For
polar semiconductors, such as Al$_x$Ga$_{1-x}$N, there exist both
acoustic and optical phonon modes. For optical phonon modes, only
the longitudinal-optical phonon mode can strongly couple to
electrons. For acoustic phonon scattering, on the other hand, we
use the deformation-potential approximation \cite{lyo2} with
parameters given in the text. The detailed form of electron-phonon
coupling strength $|C^{\rm e}_{q\lambda}|^2$ can been found from
publications \cite{lyo2,aposto}. Applying the Debye model to
low-energy acoustic phonons, we get
$\omega_{q\lambda}=c_{\lambda}q$ with $\lambda=\ell,\ t$. For
holes we get the similar power-exchange density ${\cal W}^{\rm
h}_{\rm s}$ with $T_{\rm h}$. It is clear from Eq.\,(\ref{e3})
that the electron energy loss or gain from phonons under weak pump
field depends on whether the electron temperature is higher or
lower than the lattice temperature, respectively. The same
argument applies to holes.
\medskip

In order to cool the lattice, the power gain of the electrons due
to laser pumping must be smaller than the power loss due to
spontaneous photon emission. This requires a very weak laser field
and a large bandgap. The energy conservation in steady state
requires

\begin{equation}
{\cal W}_{\rm ab}-{\cal W}_{\rm sp}+{\cal W}^{\rm e}_{\rm s}+{\cal
W}^{\rm h}_{\rm s}=0\ . \label{e200}
\end{equation}
The solution of this equation provides the carrier temperature for
any given lattice temperature $T_{\rm L}$ since ${\cal W}^{\rm
e}_{\rm s}+{\cal W}^{\rm h}_{\rm s}$ explicitly depend on the
lattice temperature $T_{\rm L}$. The sign of ${\cal W}_{\rm
ab}-{\cal W}_{\rm sp}$ determines the signs of $T_{\rm e}-T_{\rm
L}$ and $T_{\rm h}-T_{\rm L}$. The larger the value of $|{\cal
W}_{\rm ab}-{\cal W}_{\rm sp}|$, the larger the deviation of
carrier temperature from $T_{\rm L}$ will be.
\medskip

Although the phonons also stay in a quasi-equilibrium state, the
phonon temperature $T_{\rm L}$ directly evolves with time due to
an imbalance between the power loss to electrons and holes and the
power gain from any external thermal source (such as the
background thermal radiation). As a result, the average phonon
energy varies with time. This gives rise to

\[
\frac{\hbar^2}{8\pi^2k_{\rm B}T^2_{\rm L}}\ \left(\frac{dT_{\rm
L}}{dt}\right)\ \sum_{\lambda}\ \int_0^{\pi/a_{\rm L}}\ dq\ q^2\
\omega^2_{q,\lambda}\
\sinh^{-2}\left(\frac{\hbar\omega_{q\lambda}}{2k_{\rm B}T_{\rm
L}}\right)
\]
\begin{equation}
=\frac{\sigma{\cal A}_{\rm s}}{{\cal V}}\ \left(T^4_{\rm
B}-T^4_{\rm L}\right)-\left({\cal W}_{\rm sp}-{\cal W}_{\rm
ab}\right)\ , \label{e5}
\end{equation}
where $a_{\rm L}$ is the lattice constant, $\sigma=\pi^2k_{\rm
B}^4/60\hbar^3c^2$ is the Stefan-Boltzmann constant, ${\cal
A}_{\rm s}$ is the surface area of the sample. We assume $T_{\rm
L}=T_0$ at $t=0$, where $T_0$ is the initial temperature of
equilibrium phonons, and $T_{\rm B}$ is the environmental
temperature which is close to $T_0$ for bandedge pumping with very
weak laser field. The first term in Eq.\,(\ref{e5}) is much
smaller than the second term even when $T_{\rm L}\neq T_{\rm B}$.
The rate of reduction of $T_{\rm L}$ is determined by ${\cal
W}^{\rm e}_{\rm s}+{\cal W}^{\rm h}_{\rm s}$ which decreases with
decreasing $T_{\rm L}$ and the temperature difference $T_{\rm
L}-T_{\rm e}$. Moreover, we note that the bandgap of
semiconductors in general depends on the lattice temperature, but
can be neglected for wide bandgap semiconductors such as
Al$_x$Ga$_{1-x}$N. From Eq.\,(\ref{e5}) we know that the cooling
of the lattice implies ${\cal W}^{\rm e}_{\rm s}+{\cal W}^{\rm
h}_{\rm s}={\cal W}_{\rm sp}-{\cal W}_{\rm ab}>0$. This requires
$T_{\rm e}<T_{\rm L}$ ($T_{\rm h}<T_{\rm L}$ for hole) from
Eq.\,(\ref{e3}) with a weak pump field and a large bandgap.
\medskip

In this paper, we consider the semiconductor $Al_xGa_{1-x}N$ for
our numerical calculations, where $x$ is the percentage of $Al$ in
the alloy. The bandgap increases with $x$.
\begin{figure}[h]
\begin{center}
\includegraphics[width=5.0in,height=4.0in]{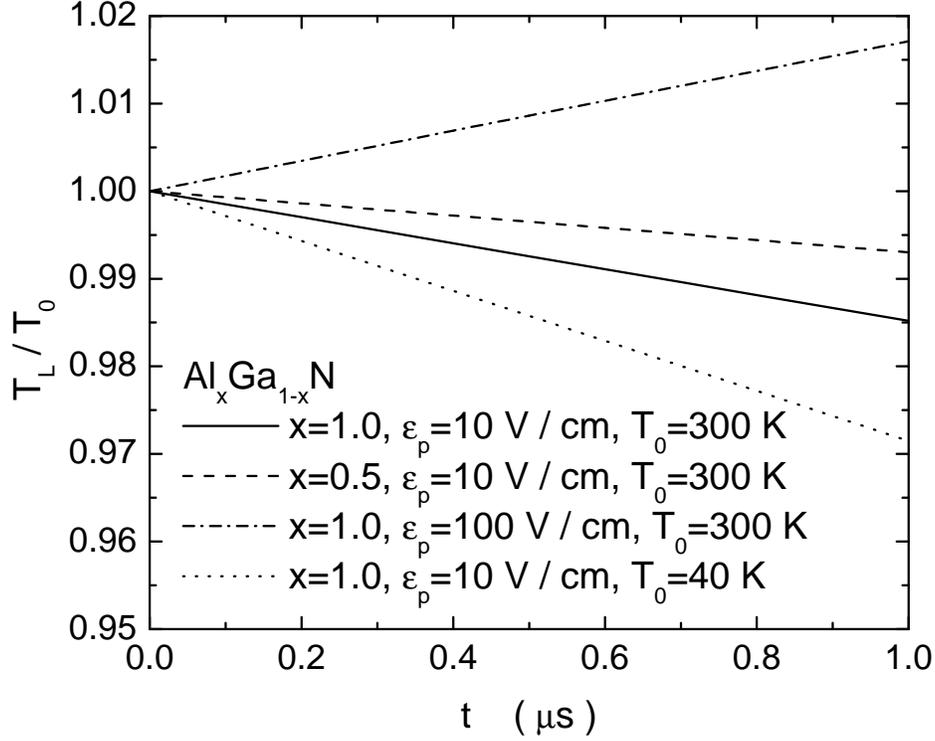}
\end{center}
\caption{Calculated scaled lattice temperature $T_{\rm L}/T_0$ for
Al$_x$Ga$_{1-x}$N as a function of time $t$ with $\hbar\Omega_{\rm
p}-E_{\rm G}=10\ meV$ for four different cases. These cases
include: $x=1$ and ${\cal E}_{\rm p}=10\ V/cm$, $T_0=300$ K (solid
curve); $x=0.5$ and ${\cal E}_{\rm p}=10\ V/cm$, $T_0=300$ K
(dashed curve); $x=1$ and ${\cal E}_{\rm p}=100\ V/cm$, $T_0=300$
K (dash-dotted curve); and $x=1$, ${\cal E}_{\rm p}=10\ V/cm$,
$T_0=40$ K (dotted curve). The other parameters are given in the
text.} \label{f1}
\end{figure}

For $Al_xGa_{1-x}N$, we choose the following parameters: $E_{\rm
G}=3.4+2.8x\ eV$, $m_{\rm e}^{\ast}/m_0=0.2+0.2x$, $m_{\rm
h}^{\ast}/m_0=1.4+2.13x$, $\Delta_0=0.02-0.001x\ eV$,
$\epsilon_{\rm s}=8.9-0.4x$, $\epsilon_{\infty}=5.35-0.58x$,
$\hbar\omega_{\rm LO}=91.2+8.0x\ meV$, $\rho=6.15-2.92x\ g/cm^3$,
$D=-(8.3+1.2x)\ eV$, $h_{14}=(2.81+4.09x)\times 10^7\ V/cm$,
$c_t=(2.68+1.02x)\times 10^5\ cm/sec$,
$c_{\ell}=(6.56+2.56x)\times 10^5\ cm/sec$, $a_{\rm
L}=5.12-0.14x$, $\gamma_0=\hbar/\tau$ with $\tau=0.1\ ps$, $n_{\rm
i}=10^{10}\ cm^{-3}$, $\epsilon_{\rm r}=(\epsilon_{\rm
s}+\epsilon_{\infty})/2$, $\hbar\Omega_{\rm p}-E_{\rm G}=10\ meV$,
and the sample is assumed to be cubic with an edge size of $1\
cm$.
\medskip

Figure\,\ref{f1} displays our main results for the scaled lattice
temperature $T_{\rm L}/T_0$ as a function of time $t$ ($0\leq
t\leq 1$ $\mu$s) for Al$_x$Ga$_{1-x}$N. From it we find that the
laser cooling at $T_0=300\ K$, $x=1.0$ and ${\cal E}_{\rm p}=10\
V/ cm$ (solid curve) is the largest compared to the other three
cases, reaching as high as $k_{\rm B}\Delta T_{\rm L}/\Delta
t=k_{\rm B}(T_0-T_{\rm L})/\Delta t=380$ eV$/$s.
The cooling effect becomes smaller when $x$ is reduced to $0.5$ (dashed curve)
with a smaller bandgap. Moreover, the laser cooling changes into
laser heating when ${\cal E}_{\rm p}$ is increased to $100$ V$/$cm
(dash-dotted curve). Finally, the laser-cooling efficiency
decreases to $100$ eV$/$s when $T_0$ drops to $40$ K (dotted
curve). This indicates that the laser cooling of a lattice can be
maximized for wide-bandgap semiconductors under the conditions of
low pump-field strength and high initial lattice temperature.
\medskip

In conclusion, by using the energy-balance equation for pump-laser
induced conduction electrons and holes, we have demonstrated a
laser-cooling power as high as $380$ eV$/$s for the wide bandgap
semiconductor AlN at room temperature when the pump-laser field is
only $10$ V$/$cm. The evolution of the lattice temperature was
found to be determined by the difference between the power-loss
density due to photoluminescence and the power-gain density due to
optical absorption, as well as the initial lattice temperature.
\medskip

The authors are grateful for helpful discussions with Prof. M.
Sheik-Bahae from the University of New Mexico, Dr. B. Flake from
the Air Force Research Lab, and Dr. R. I. Epstein from the Los
Alamos National Laboratories.

\end{document}